\documentclass[twocolumn,showpacs,preprintnumbers,amsmath,amssymb]{revtex4}
\usepackage{graphicx}
\usepackage{color}
\newcommand {\bea} {\begin{eqnarray}}
\newcommand {\eea}{\end{eqnarray}}
\begin{document}
\draft
\title{Equation of state of a polarized Fermi gas in the Bose-Einstein Condensate limit}
\author{F. Alzetto and X. Leyronas}
\affiliation{
Laboratoire de Physique Statistique, Ecole Normale Sup\'erieure, UPMC
Paris 06, Universit\'e Paris Diderot, CNRS, 24 rue Lhomond, 75005 Paris,
France.}

\begin{abstract}
We present a theoretical study of the BEC-BCS crossover in the Bose-Einstein-Condensate regime (BEC), in the case of an unequal number of fermions
of two species. We take full account of the composite nature of the dimers made of fermions. In the limit of low densities, we calculate the ground state energy of the system, or equivalentely the chemical potentials of each species as well as the one-particle gap and the energy of an "impurity" immersed in a Fermi sea. For the chemical potentials we go up to order $(density)^{4/3}$.The results found involve the exact atom-dimer $a_{AD}$ and dimer-dimer $a_{DD}$ scattering lengths and therefore
include the $3$ and $4$-body problems in the manybody problem. We briefly comment
on the importance of the different mean-field corrections for recent experiments.

\end{abstract}

\maketitle

\section{Introduction}
The problem of a Bose-Fermi mixture in the Bose-Einstein-Condensate (BEC) limit of the
BEC-BCS (Bardeen-Cooper-Schrieffer) crossover has been studied experimentally 
recently \cite{YShinKetterle}. Previous theoretical works studied the zero temperature phase diagram using a mean field approach \cite{becbcsmf}, or Monte Carlo simulations \cite{pilatigiorgini}. The BEC limit has been investigated theoretically in Refs.\cite{pieristrinati},\cite{iskinsademelo} or using a functional integral formalism  \cite{edtaylor}. In Ref.\cite{pieristrinati}, the authors used a Bogoliubov-de Gennes approach in order to derive in an elegant way a Gross-Pitaevskii
equation for the condensate wave function. They found a dimer-dimer scattering length
and an atom-dimer scattering length equal to $2$ times the atom-atom scattering length
$a$ and $8/3\,a$ respectively. Then they gave some reasonable diagrammatic arguments in order to recover the correct scattering lengths. As far as the equation of
state of the system is concerned, these calculations stopped at the mean-field level
(where the chemical potentials depend linearly on the densities).
On the other hand, researchers had studied previously \cite{albusetal},\cite{vivgior}
a mixture of point-like bosons and fermions at zero temperature in the dilute limit.
There the equation of state was determined including a lowest order mean-field correction. This is the main result of the present work to recover this result for the
problem of a mixture of fermions and composite bosons relevant to experiment 
of Ref.\cite{YShinKetterle}. 

In this work we determine the equation of state at zero temperature of a mixture of fermions of two species. We consider a {\it dilute} situation where the densities of both species $n_1$ and $n_2$ (hereafter denoted $density$) are small but are of the same order.
We thus have $density\,\times\,a^3\,\ll 1$, and we expand the chemical potentials of fermions up to order $(density)^{4/3}$.

For a reader mainly interested in the results, and not so much in technical details, the main results of our work (the equations of state) are shown in Eqs.\ref{eqmu1f},\ref{eqmu2} and \ref{eqmubf}.

In Sec.\ref{general}, we recall the general formalism of fermionic superfluids (anomalous Green's functions, anomalous self-energy, and the so-called generalized self-energies).
In Sec.\ref{gapnb}, we describe the general framework of the theory, namely how we expand the number and gap equations in powers of the atomic density. In Sec.\ref{densexp1}, the lowest order density expansion is deduced, while the derivation of the next order correction is shown in Sec.\ref{densepx2}. The one-particle gap and the problem of a single atom immersed in the Fermi Sea of atoms of the other species are determined in Sec.\ref{gapimp}.
In Sec.\ref{conclusion}, we conclude and comment on the relevance of our calculation for recent experiments of Ref.\cite{YShinKetterle}.
\section{General formalism}\label{general}
\begin{figure}[h]
\begin{center}
\includegraphics[width=8cm]{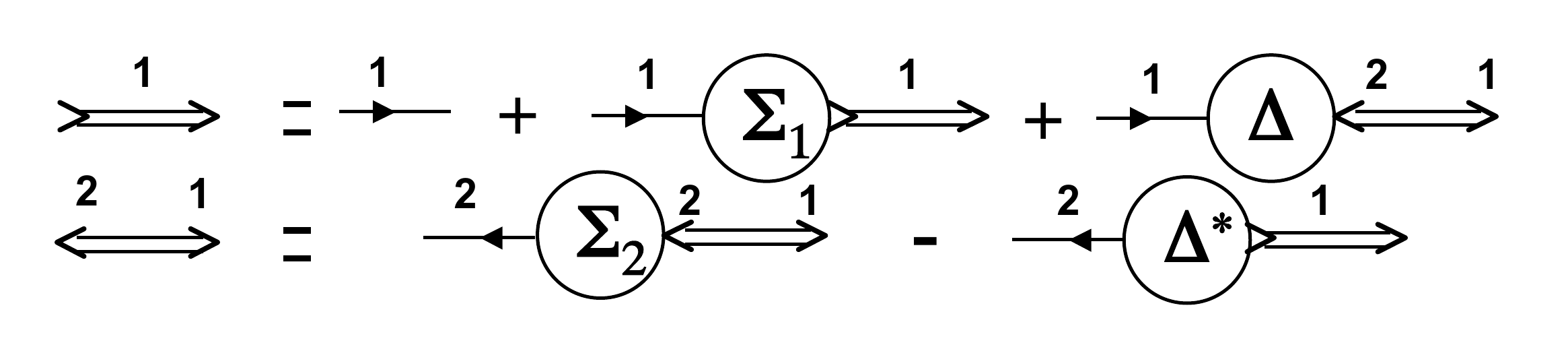}
\caption{The Dyson equations Eq.\ref{eqDyson1} and \ref{eqDyson2} in diagrammatic language. The numbers $1$ and $2$ specify the atomic species.}
\label{figDyson}
\end{center}
\end{figure}

In a fermionic superfluid, the usual field-theoretical methods must be modified in order to take into account the off-diagonal order \cite{AGD}. In practice, one has to consider
"anomalous" Green's functions  $F(p)$ and $F^{+}(p)$ (which are physically the propagators of condensed pairs). The Dyson equations for the fermionic superfluid are
\bea
G_1(p)&=&G_1^0(p)+G_1^0(p)\Sigma_1(p)G_1(p)+G_1^0(p)\Delta(p) F^{+}(p)\nonumber\\\label{eqDyson1}
\eea
for the propagator of particles of type $1$ (and similarly for particles of type $2$) and
\bea
F^{+}(p)&=&G_2^0(-p)\Sigma_2(-p)F^{+}(p)-G_2^0(-p)\Delta^*(p)G_1(p)\nonumber\\\label{eqDyson2}
\eea
for the anomalous Green's function $F^{+}(p)$.
These two equations have a simple diagrammatical meaning (see Fig.\ref{figDyson}):
the self-energy $\Sigma_1(p)$ is the sum of all diagrams with one incoming $(p,1)$ line (a $G_{1}^0(p)$ Green's function) and one outgoing $(p,1)$ line, which can not be separated into two disconnected parts by cutting a $(p,1)$ line or a $(-p,2)$ line
(a $G_{2}^0(-p)$ Green's function). The anomalous self-energy $\Delta(p)$ is the sum of all diagrams with one incoming $(p,1)$ line and one incoming $(-p,2)$ line, which can not be separated into two disconnected parts by cutting a $(p,1)$ line or a $(-p,2)$ line.
The anomalous Green's function $F^{+}(p)$ starts diagrammatically with one outgoing $(-p,2)$ line and finishes with one outgoing $(p,1)$ line. 

In Ref.\cite{xlrclhy}, the so-called generalized self-energy $S_1(p)$  was introduced. It is defined by the equation
\bea
G_1(p)&=&\frac{1}{(G^{0}_{1}(p))^{-1}-S_1(p)}
\eea
which can be rewritten
$G_1(p)=G^{0}_{1}(p)+G^{0}_{1}(p)S_1(p)G^{0}_{1}(p)+G^{0}_{1}(p)S_1(p)G^{0}_{1}(p)S_1(p)G^{0}_{1}(p)+\cdots$. Diagrammatically, this means that $S_1(p)$ is the sum of all diagrams with one incoming $(p,1)$ line and one outgoing $(p,1)$ line which cannot be disconnected into two parts by cutting a $(p,1)$ line. It is natural to use the generalized self energy, since it is the only thing one needs in order to calculate the one particle Green's function
$G_1(p)$ (whereas one needs both $\Sigma_1(p)$ {\it and} $\Delta(p)$) and therefore the density of fermions of species $1$. Moreover, as it was shown in Ref.\cite{xlrclhy}, the generalized self-energy is directly related to the dimer-atom $T$-matrix (hereafter denoted $T_3$, as we will see in Section \ref{densexp1}), whereas $\Sigma_1(p)$ is not. This makes the use of the generalized self-energy convenient to get more compact expressions.
\section{Number and gap equations}\label{gapnb}
The system we study is an ensemble of two distinct "species" of fermions. There are 
$n_1$ (respectively $n_2$) fermions per unit volume of type $1$ and mass $m_1$ (resp. type $2$ and mass $m_2$). These two species can be either two internal states (for instance
 two hyperfine states of $^{6}Li$) and in this case $m_1=m_2$ or two different sort of
 fermions (a $^{6}Li$-$^{40}K$ mixture for instance). Due to the extremely low temperature and Pauli exclusion principle there is no intra-species interaction. The short range inter-species interaction is therefore described by the scattering length $a$
 (for a review, see Ref.\cite{gps}).
 In the limit we consider, we have $a>0$ and there is a two-body bound state (a molecule or dimer) of energy $E_b=-1/(2 m_r a^2)$ where $m_r= m_1 m_2 /(m_1+m_2)$
 is the particle $1$-particle $2$ reduced mass and we will take in this work  \mbox{$\hbar=1$}.
  As pointed out by Pieri and Strinati \cite{pieristrinati}, the limit of dilute condensed molecules with
extra fermions of species $1$ corresponds to a positive chemical potential $\mu_1$ 
and a negative
and large chemical potential (of order $E_b$) $\mu_2$ for the fermions $2$. 
Therefore this is the limit we will consider too and we take
\bea
\mu_1&>&0\\
\mu_2&<&0
\eea
and 
\bea
\mu_1+\mu_2&<&0
\eea
 The method is similar to the one used in Ref.\cite{xlrclhy} and \cite{praclk}, and is common to most of the studies of the BCS-BEC crossover \cite{gps} :
 we  calculate the particles number $n_1$ and $n_2$ as a function of the chemical potentials $\mu_1$ and $\mu_2$ and a superfluid order parameter $\Delta$; we write down also
 a generalized "gap equation" for $\Delta$ as a function of $\mu_1$ and $\mu_2$.
 If we eliminate $\Delta$ from the equations, we get the equations of state
 $\mu_1(n_1,n_2)$ and $\mu_2(n_1,n_2)$. The results for these equations of state
 are shown in Eqs.\ref{eqmu1f},\ref{eqmu2} and \ref{eqmubf}, which are the main results of the present work.
\subsection{Number equations}
In order to get the density of particle $1$, we have to sum on $k$ the propagator 
$G_1(k)$
\bea
n_1&=&-\sum_k G_1(k)e^{i\omega_n 0^+}
\eea
where we use the notation $\sum_k\equiv -\int \frac{d^{3}{\bf k}}{(2\pi)^3} T\sum_{n}$
(in all the paper, we use the $T\neq 0$ formalism and then take the limit $T\to 0$) . 
We have  $G^0_{1}(k)=\frac{1}{\left(i\omega_n+\mu_1-\frac{{\bf k}^2}{ 2\,m_1}\right)}$ with $k=\{{\bf k},i\omega_n\}$, $\mu_1>0$ and $\omega_n=(2n+1)\pi T$ (with $n$ integer). We have of course a similar result for the Green's function of atoms $2$, with a
chemical potential $\mu_2 <0$. 
Following \cite{xlrclhy}, the full Green's function $G_1(k)$ is related to the generalized
self-energy $S_1(k)$ via a Dyson equation $G_1(k)=G^0_1(k)+G^0_1(k) S_1(k) G_1(k)$. $S_1(k)$ does not contain diagrams with a $G^0_1(k)$ propagator, but it does contains diagrams with $G^0_2(-k)$ propagators. If we iterate the Dyson equation, we get the approximation
\bea
G_1(k)&\approx G^0_1(k)+(G^0_1(k))^2 S_1(k)\label{approxG1}
\eea
As we explain below, this approximation is good enough at the order we consider.
It must be emphasized that the generalized self-energy contains anomalous contributions, as we will see.
Using the approximation Eq.\ref{approxG1}, we get from the first term the contribution
to the density (after taking the $T\to 0$ limit) 
\bea n^0_1&=&\frac{k_{0}^3}{6\pi^2}\label{eqn10}
\eea
where the wave-vector $k_0$ is defined
via 
\bea
k_0&=&\sqrt{2m_1 \mu_1}
\eea 
If we define $\delta n_1 \equiv n_1 -n^0_1$, we get
\bea
\delta n_1&\approx&-\sum_k (G^0_1(k))^2 S_1(k)
\eea
 We can proceed in the same way for the density of particles $2$. In this case
 we have $n^0_2 = 0$, since $\mu_2 <0$. We therefore have
 \bea
n_2&\approx&-\sum_k (G^0_2(k))^2 S_2(k)\label{eqn2sum}
\eea
where $S_2(k)$ is the generalized self-energy of the particles $2$.
\subsection{Gap equation}
  We follow Ref.\cite{xlrclhy} in order to get our gap equation. Looking at the diagrammatic expansion of the anomalous Green's function $F(k)$, we can in general write :
\bea
F(k)&=&G^0_{1}(k)G^0_{2}(-k)\left(\sum_{k_1}V({\bf k}-{\bf k}_1)F(k_1)\right)\nonumber\\
 &&+F'(k)\label{eqgap1}
\eea
The first term in the r.h.s. of Eq.\ref{eqgap1} contains {\it all} the diagrams where the two incoming $(k,1)$ and $(-k,2)$ fermions lines first interact (interaction $V({\bf q})$).
Therefore $F'(k)$ is the sum of all the diagrams where the two incoming fermions
$(k,1)$ and $(-k,2)$ do {\it not} first interact. 
The term between parentheses in Eq.\ref{eqgap1} is the usual BCS gap \cite{xlrclhy}
$\delta_{1}({\bf k})$
\bea
\delta_1({\bf k})&=&\sum_{k_1}V({\bf k}-{\bf k}_1)F(k_1)
\eea
We can eliminate the interaction $V$ in favor of the $2$-body $T$-matrix  \cite{xlrclhy}
at zero momentum $T_2(0)$. 
\bea
\left[T_2(0)\right]^{-1}&=&\sum_k F'(k)/\Delta
\eea
where $\Delta$ is the zero wave vector value of $\delta_1({\bf k})$.
As usual, $T_2$ is obtained by summing the ladder series and one finds easily
\bea
T_2(K)^{-1}&=& g^{-1}-\Pi(K)\nonumber\\
&=&\frac{m_r}{2\pi}\left(
a^{-1}-\sqrt{-2m_r(i\Omega_{\nu}+2\mu-\frac{{\bf K}^2}{2\,m_D})} 
\right)\nonumber\\
&&+R(K)\label{eqT2}
\eea
where $K=\{{\bf K},i\omega_{\nu}\}$, $\omega_{\nu}=2\pi\nu T$ ($\nu$ integer). $m_D=m_1+m_2$ is the mass of a dimer and $2\mu=\mu_1+\mu_2$ is the total chemical potential.
 $R(K)$ is due to the positive chemical potential $\mu_1$ :
 \bea
R(K)&=& \int_{k<k_0}\!\frac{d^3 {\bf k}}{(2\pi)^3}
\frac{1}{\left( i\Omega_{\nu}+2\mu -\frac{{\bf k}^2}{ 2\,m_1}-\frac{({\bf K- k})^2}{ 2\,m_2}    \right)}\nonumber\\
\eea
 This is due physically to the presence of the Fermi Sea of the fermions in excess.
 We note that $T_2(K)$, is analytical for $\mathcal{R}e(i\omega_{\nu})<0$
 (in many instances in this work, we find convenient to use the somewhat loose notation $g(i\omega_{\nu})$ for the analytical
 continuation of a function $g$ for $i\omega_{\nu}$ spanning the whole complex plane
 - not only the discrete Matsubara frequencies values).
 The final gap equation reads 
 \bea
 \frac{m_r}{2\pi}\left(a^{-1}-\sqrt{4m_r |\mu|}\right )&=&(\sum_k F'(k)/\Delta)-R(0)\label{eqgapbase}
  \eea
  with $R(0)= -\int_{k<k_0}\frac{d^3 {\bf k}}{(2\pi)^3}
\frac{1}{\left( \frac{{\bf k}^2}{ 2 m_r}  +2|\mu| \right)}$

We {\it define} the bosonic chemical potential $\mu_B$ through the equation
\bea
\mu_1 +\mu_2 \equiv 2\mu &\equiv&-|E_b|+\mu_B\label{eqdefmuB}
\eea
and we can expand the l.h.s. of Eq.\ref{eqgapbase} in power of $\mu_B$. In this way we get
\bea
\mu_B &=& Z \left[(\sum_k F'(k)/\Delta)-R(0)\right]+O(\mu_{B}^2)\label{eqgapmub}
\eea
where we have defined the residue of $T_2$ at the molecular pole
\bea
Z &=& \frac{2\pi}{(m_r)^2 a} \label{eqdefZ}
\eea
In the following, we will expand $F'(k)$, $S_1(k)$ and $S_2(k)$ in power of $\Delta$ and $k_0$. We will see this corresponds to an expansion in power of the density of the condensed dimer and
of the fermions in excess respectively.

\section{Density expansion : lowest order}\label{densexp1}
\subsection{Number equations}
\begin{figure}[h]
\begin{center}
\includegraphics[width=8cm]{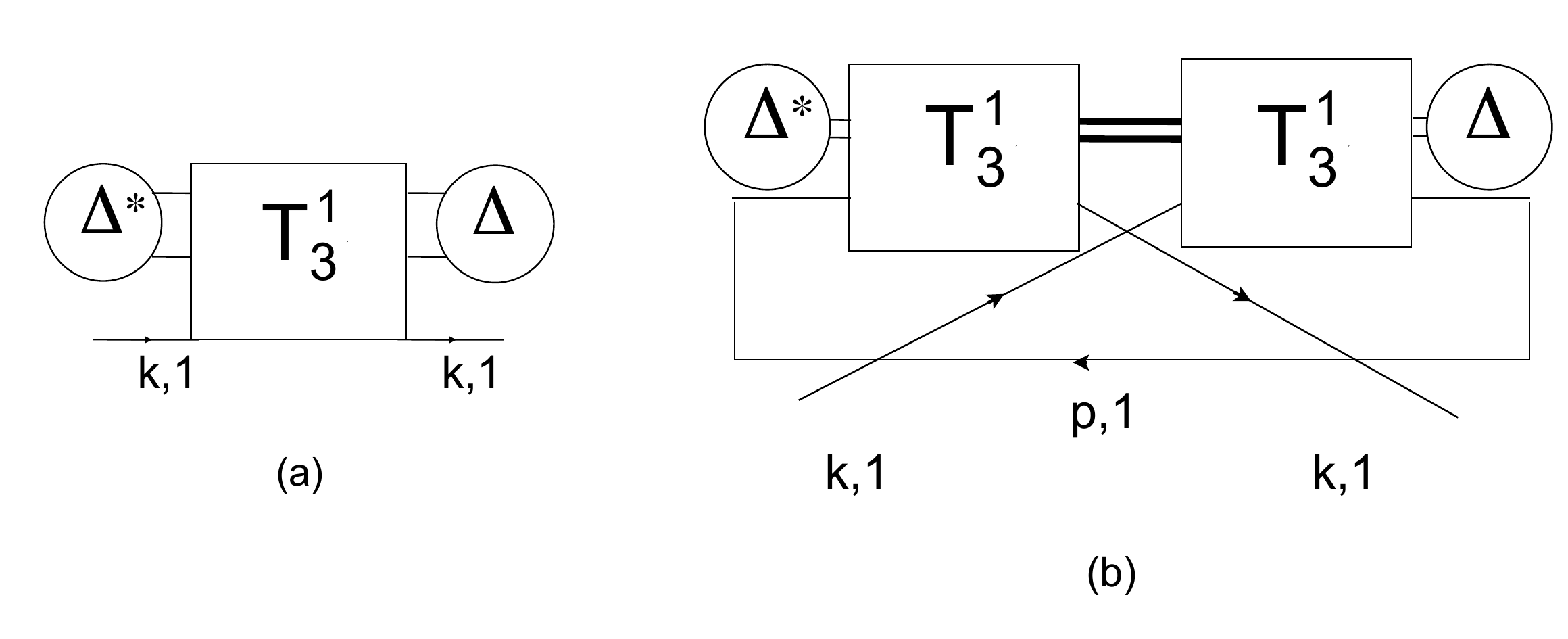}
\caption{Diagrammatic representation of $S_{1}^{(a)}(k)$ and $S_{1}^{(b)}(k)$.}
\label{fig2}
\end{center}
\end{figure}

We start with the calculation of $S_1(k)$. Diagrams without anomalous self-energy are diagrams in the normal phase. In this case, one has to close a fermionic loop for the species $2$. Going back to time representation, one sees that such contributions are zero, since the bare propagators of particles $2$ are retarded, as is explained in details in \cite{xlrclhy}. Physically, this comes from the fact that at $T=0$, there are no "free" ({\it i.e} not bound to a condensed molecule) fermion of species $2$ which can interact with a fermion of species $1$. The diagrams of lowest order in $\Delta$   (that is of lowest
order in $\sqrt{n_2}$) are therefore those containing one $\Delta$ and one $\Delta^*$. They
are given in terms of the $3$-particle $T$-matrix $T_3^1$ describing the scattering of a dimer with
fermion of species $1$ (see Fig.\ref{fig2}(a))
\bea
S_1^{(a)}(k)&=&|\Delta|^2 T_3^{1}(k,k;k)\label{eqS1a}
\eea
This diagram has a very simple physical meaning. An incoming fermion $k,1$ scatters with a condensed dimer (factor $\Delta^*$ for a dimer "out of" the condensate), the scattering of the dimer with the fermion is described by the dimer-atom $1$ $T$-matrix $T_3^1$. The right of the graph represents the outgoing fermion $k,1$ and the dimer which "goes back" to the condensate (factor $\Delta$).
The contribution to the density of particles $1$ is 
$\delta n_1^{(a)}=-\sum_k e^{i\omega_n 0^+} (G_1^0(k))^2 |\Delta|^2 T_3^{1}(k,k;k)$.
The frequency summation is written as usual as an integral on a contour encircling the imaginary axis counterclockwise :
$
T\sum_{i\omega_n} (G_1^0(k))^2 T_3^{1}(k,k;k)
=
-\oint_{\mathcal{ C}}\frac{dz}{2\pi i} f(z) \frac{1}{(z-\xi_1 )^2} T_3^{1}(k,k;k)
$
with $f(z)=1/(e^{z/T}+1)$ is the Fermi Dirac distribution and $\xi_{1,2}={\bf k}^2/(2 m_{1,2})
-\mu_{1,2}$ and $k=\{{\bf k},z\}$.
 We use next the analytical properties of $T_3$ as a function of $z$.
Following an argument given in \cite{xlrclhy}, we can split $T_3$ in a first part
$1/(z+\xi_2)$ which has a pole at $-\xi_2<0$ plus a function which is analytical for
$\mathcal{R}e(z)<0$.
When we deform the contour in order to enclose the singularities, we see that the only singularities in $\mathcal{R}e(z)<0$ are $-\xi_2$ and $\xi_1$ if $\xi_1<0$. Due to the exponential decay of the Fermi-Dirac distribution $f(z)$ for $\mathcal{R}e(z)<0$ when the temperature goes to $0$, the
contour encircling the domain $\mathcal{R}e(z)>0$ vanishes.
We therefore find at zero temperature

\bea
T\sum_{i\omega_n} G_1^0(k)^2 T_3^{1}(k,k;k)
&=&
\frac{f(-\xi_2)}{(\xi_1+\xi_2)^2}+\frac{d f(\xi_1)}{d z} T_3^{1}(k,k;k)\nonumber\\
&&+f(\xi_1)\frac{d}{dz} T_3^{1}(k,k;k)\label{eqdnk1a}
\eea
where the $4$-wave vector $k=\{{\bf k},\xi_1\}$ is "on the shell". In order to compute the density the next step is to integrate the three terms on the {\it r.h.s.} of Eq.\ref{eqdnk1a} on the wave vector ${\bf k}$.  We use the zero temperature limits $\frac{d f(\xi_1)}{d z}=-\delta(\xi_1)$, $f(\xi_1)=\Theta(k_0-|{\bf k}|)$ 
($\Theta(x)$ is the step function) and $f(-\xi_2)=1$.
The first term is simply $\int\!\frac{d^3 {\bf k}}{(2\pi)^3} \frac{1}{(\frac{{\bf k}^2}{2 m_r}+2|\mu|)^2} $. However, anticipating the result, we can replace $2\mu$ by the binding energy $E_b=-\frac{1}{2 m_r a^2}$ plus terms of order $(density)$. We find in this way that the contribution to $n_1/|\Delta|^2$ of the first term of Eq.\ref{eqdnk1a} after summation on ${\bf k}$ is 
\bea
\frac{(m_r)^2 a}{2\pi}=Z^{-1}\label{eq1term}
\eea

For the second term of Eq.\ref{eqdnk1a}, due to the $\delta$ function,  the integration 
gives a factor $N_0=m_1 k_0/(2\pi^2)$, the density of state of excess fermions at energy $\mu_1$, times (minus) $T_3^1(k,k;k)$ at $k=\{k_0,0\}$. Then we use Eq.\ref{eqdt3f} of Appendix \ref{appdt31} in order to find $T_3^1$ up to order $k_0$. In this way the contribution to $\delta n_1^{(a)}/|\Delta|^2$ of the second term  after summation on ${\bf k}$ is
\bea
-Z^{-1}N_0 g_{AD}[1+k_0 a_{AD}\,(2\pi)\,F_{a,1}(1,m_D/m_1)]\label{eq2term}
\eea
where we have introduced the dimer-particle $1$ coupling strength $g_{AD}=2\pi\, a_{AD}/m_{AD}$, where $m_{AD}=(m_1 m_D)/(m_1+m_D)$ is the dimer-particle $1$ reduced mass and $a_{AD}$ is the dimer-fermion $1$ scattering length. The function $F_{a,1}$ is defined in Eq.\ref{eqFa1} of Appendix \ref{appdt31}.
Concerning the third term on the {\it r.h.s.} of Eq.\ref{eqdnk1a}, we use the result of Eq.\ref{eqddt3f} of Appendix \ref{appddt31} ($k=\{{\bf k},z=({\bf k}^2 -k_0^2)/2m_1\}$)
\bea
\frac{d}{dz} T_3^{1}(k,k;k)&=&-Z^{-1}(g_{AD}m_{AD})^2\frac{1}{k_0}F_{a,2}\left(\frac{|{\bf k}|}{k_0},\frac{m_D}{m_1}\right)\nonumber
\eea
The function $F_{a,2}$ is defined in Eq.\ref{eqFa2} of Appendix \ref{appddt31}.
After integration on ${\bf k}$, due to the occupation number $f(\xi_1)=\Theta(k_0-|{\bf k}|)$,  we find
$
\int_{|{\bf k}|<k_0}\!\frac{d^3 {\bf k}}{(2\pi)^3}
\frac{d}{dz} T_3^{1}(k,k;k)
$.
If we rescale ${\bf k}$ by $k_0$ we find the contribution to $n_1/|\Delta|^2$ for the third term of Eq.\ref{eqdnk1a}
\bea
-Z^{-1}g_{AD} N_0 k_0 a_{AD}
\frac{m_{AD}}{m_1}(4\pi^3)\int_{x<1}\!\frac{d^3 {\bf x}}{(2\pi)^3} F_{a,2}\left(x,\frac{m_D}{m_1}\right)\nonumber\\
\label{eq3term}
\eea
We just have to multiply the three contributions of Eqs.\ref{eq1term},\ref{eq2term} and
\ref{eq3term} by $|\Delta|^2$ in order to get the correction to the density of particles $1$
\bea
\delta n_1^{(a)}&=&\frac{|\Delta|^2}{Z}\nonumber\\
&&-N_0\left(g_{AD}\frac{|\Delta|^2}{Z}\right)\left(
1+k_0 a_{AD}(C_{a,1}+C_{a,2})
\right)
\nonumber\\
\eea
 The dimensionless constants $C_{a,1}$ and $C_{a,2}$ are given by $C_{a,1}=2\pi F_{a,1}(1,m_D/m_1)$ and
$C_{a,2}=4\pi^3 (m_{AD}/m_1)\int\! \frac{d^3{\bf x}}{(2\pi)^3}\Theta(1-x)F_{a,2}(x,m_D/m_1)$.

\begin{figure}[h]
\begin{center}
\includegraphics[width=5cm]{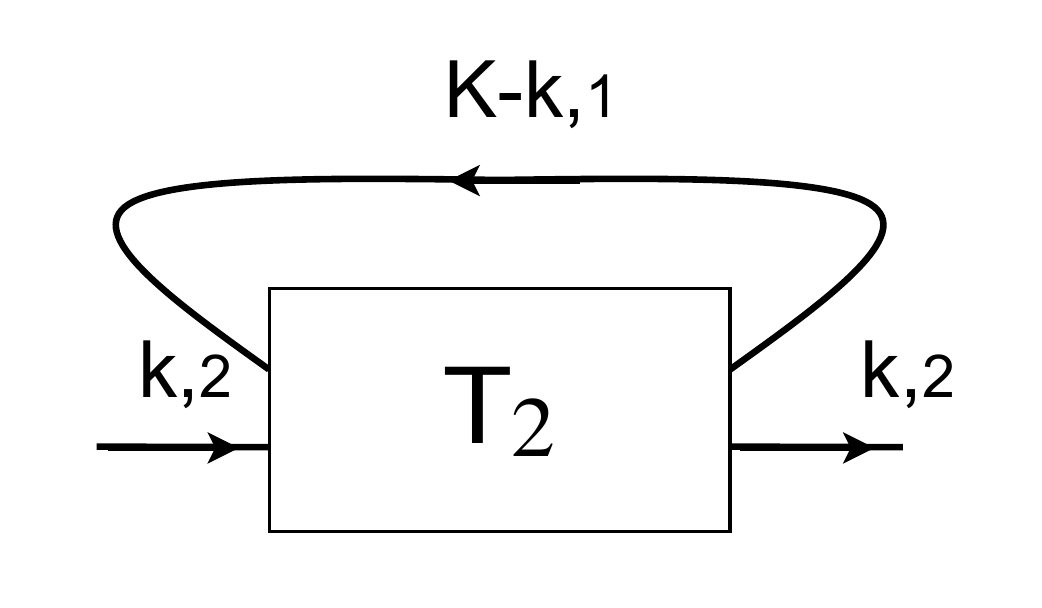}
\caption{Diagrammatic representation of $\Sigma_{2}^{(ladder)}(k)$.}
\label{figladd}
\end{center}
\end{figure}

Going to particles of species $2$, we first have to consider the self-energy of the particles $2$ in the
normal phase. The situation is different for $S_2$ and $S_1$, since, due to the positive chemical potential
$\mu_1$, the particles $2$ can interact with the Fermi Sea of particles $1$. This problem is related
to the impurity problem, which has been the object of various works both experimentally \cite{expimp} and theoretically \cite{theoimp}. In the normal phase, at lowest order in the density of particles $1$, the self-energy is given
by the so called "ladder approximation" which is shown diagrammatically  in Fig.\ref{figladd}. Explicitely it is
\bea
\Sigma_{2}^{ladder}(k)&=&-\sum_{K}T_2(K)G_1^0(K-k)
\eea
This gives the contribution to the density of particles $2$ at lowest order
\bea
n_2^{ladder}&=&-\sum_k (G_2^0(k))^2 \Sigma_{2}^{ladder}(k)\nonumber\\
&=&\sum_{k, K} (G_2^0(k))^2 T_2(K)G_1^0(K-k)
\eea
We can make the change of variables $p=K-k$ in the double summation :
$n_2^{ladder}=\sum_{k, p} (G_2^0(k))^2 T_2(k+p)G_1^0(p)$. In this way, we see that the integrand is analytical for $\mathcal{R}e(i\omega_k)<0$ and if we perform the frequency summation on $i\omega_k$, we get 
zero by deforming the frequency contour in the half-plane $\mathcal{R}e(i\omega_k)<0$. We thus find
\bea
n_2^{ladder} &=&0
\eea

We have also the contribution of the generalized self-energy due to the condensate
of dimers, similar to Eq.\ref{eqS1a} but for the particles $2$ now
\bea
S_2^{(a)}(k)&=&|\Delta|^2 T_3^{2}(k,k;k)
\eea
The contribution to the density of particles $2$ is using Eq.\ref{eqn2sum}
\bea
n_2^{(a)}&=&-\sum_k (G_2^0(k))^2 S_2^{(a)}(k)
\eea
It is shown in Appendix \ref{appn2} that at the order we consider we have
({\it i.e.} up to order $(density)^{5/3}$)
\bea
n_2^{(a)}&=&\frac{|\Delta|^2}{Z}\label{eqn2f}
\eea
Before closing this section, we notice that we could have written 
\bea
S_2(k)&\approx&\Sigma_{ladder}(k)+S_2^{(a)}(k)
\eea
and use Eq.\ref{eqn2sum} to calculate $n_2$.
\subsection{Gap equation}
\begin{figure}[h]
\begin{center}
\includegraphics[width=8cm]{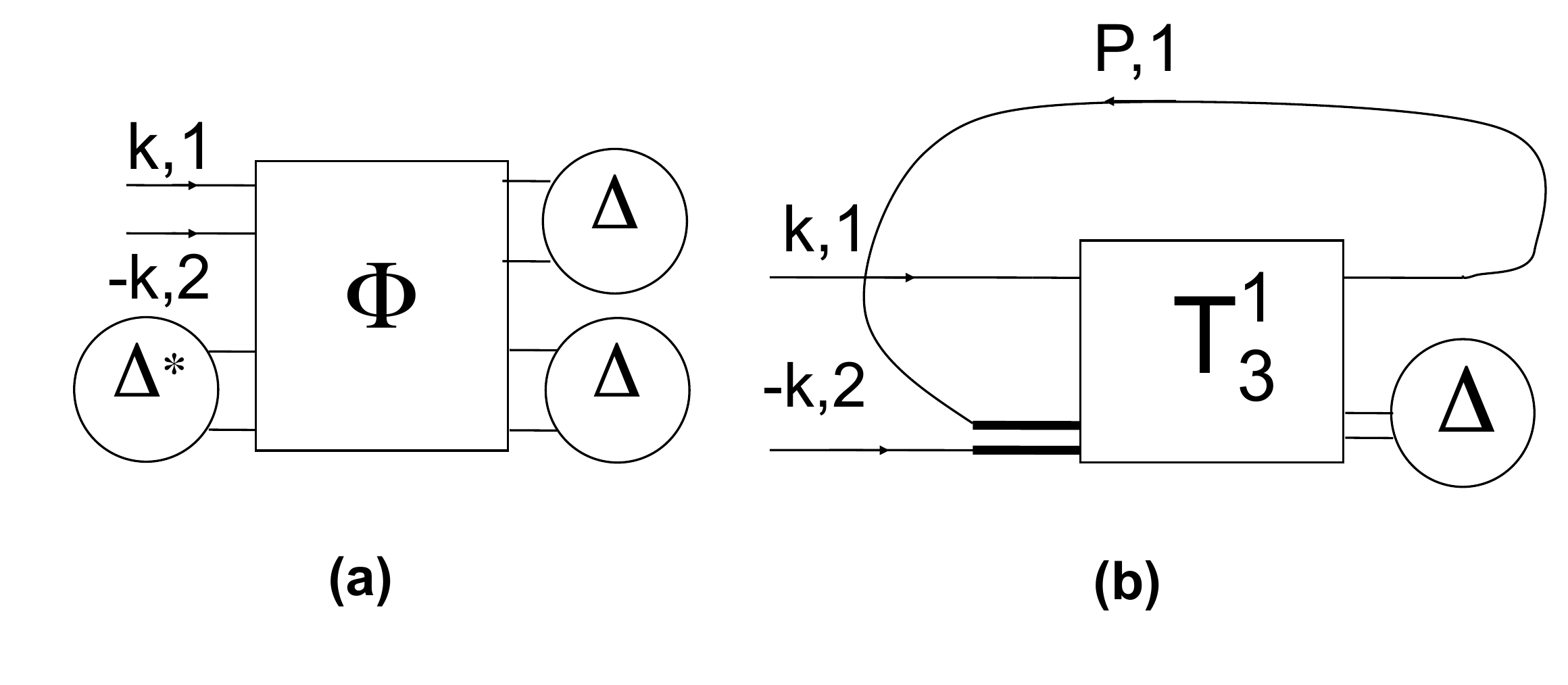}
\caption{The two contributions to $F'(k)$.}
\label{figgap}
\end{center}
\end{figure}

We determine now the lowest order contribution to $F'(k)$. We first have the
contribution already found in Ref.\cite{xlrclhy} (see Fig.\ref{figgap}(a))
\bea
\frac{F'^{(a)}(k)}{\Delta}&=&\frac{1}{2} |\Delta|^2 G_1^0(k) G_2^0(-k)\Phi(k,-k;0,0)
\eea
When $F'^{(a)}(k)$ is summed on $k$ in the gap equation \ref{eqgapmub}, we recover the dimer-dimer $T$ matrix $T_4$ :
\bea
\sum_k \frac{F'^{(a)}(k)}{\Delta}&=&\frac{1}{2}T_4(0,0;0,0)|\Delta|^2\label{eqgapT4}
\eea
However there is an other contribution to $F'(k)$ due to the presence of the Fermi Sea of the extra $1$ atoms, with one $\Delta$ and one hole propagator (see Fig.\ref{figgap}(b))
\bea
\frac{F'^{(b)}(k)}{\Delta}&=&-G_1^0(k) G_2^0(-k)\sum_p G_1^0(p) T_2(p-k) T_3^1(k,p;p)
\nonumber\\
\label{eqFb}\eea
Then $F'^{(b)}(k)$ is summed on $k$ and the order of summations on $k$
and $p$ is exchanged. The summation on $k$ (including the minus sign) is precisely the integral part in the integral equation for $T_3^1(p,p;p)$ \cite{4parpra}. Therefore we have
to substract the Born approximation term $-G_0^2(-p)$ 
$$
-\sum_k G_1^0(k) G_2^0(-k) T_2(p-k) T_3^1(k,p;p)
=T_3^1(p,p;p)+G_2^0(-p)
$$
In this way, we can transform the summation on $k$ of Eq.\ref{eqFb} in a summation on $p$
\bea
\sum_k \frac{F'^{(b)}(k)}{\Delta}&=&\sum_p G_1^0(p)[T_3^1(p,p;p)+G_2^0(-p)]
\eea
The quantity between brackets is analytical for $Re(i\omega_p)<0$. We can therefore perform the integration on $i\omega_p$ and close the contour in the half-plane
$Re(i\omega_p)<0$. We pick the contribution of the pole of $G_1^0(p)$ when $|{\bf p}|< k_0$ and we get
\bea
\sum_k \frac{F'^{(b)}(k)}{\Delta}&=&\int_{p<k_0}\frac{d^3 {\bf p}}{(2\pi)^3}
[T_3^1(p,p;p)+G_2^0(-p)]\nonumber\\
\label{eqgapFb}
\eea
where $p$ is "on the shell" $p=\{{\bf p},({\bf p}^2 -k_0^2)/2 m_1\}$.
The second term in the brackets of Eq.\ref{eqgapFb} is canceled by $-R(0)$ in the gap equation \ref{eqgapmub}.
Finally, we can write a final gap equation in terms of the bosonic chemical potential
$\mu_B$ from Eqs.\ref{eqgapmub},\ref{eqgapT4},\ref{eqgapFb}
\bea	
\mu_B &\approx& Z\left(\frac{1}{2}T_4(0,0;0,0)|\Delta|^2 +\int_{p<k_0}\frac{d^3 {\bf p}}{(2\pi)^3}
T_3^1(p,p;p)\right)\nonumber\\
\eea 
where we have omitted terms of order $\mu_B^2$, {\it i.e.} of the order of the square of 
density. We will see this is sufficient at the order we consider.

We can express $T_4$ in terms of the dimer-dimer scattering length $a_{DD}$ \cite{4parpra}. We have 
$T_4(0,0;0,0)=Z^{-2}(4\pi/m_D)(2 a_{DD})$. The dimer-dimer coupling constant is defined as usual by $g_{DD}=(4\pi a_{DD})/m_D$. Moreover, we can use the expansion of Appendix \ref{appdt31} Eq.\ref{eqdt3f} for $T_3^1(k,k;k)$ at lowest order :
$T_3^1(k,k;k)\approx g_{AD}/Z$. In this way, we can perform the integration on the wave vector ${\bf k}$ and we get the mean-field like contributions
$\mu_B=g_{DD}(|\Delta|^2 / Z)+g_{AD} (k_0^3/6\pi^2)$.
Using Eqs.\ref{eqn10} and \ref{eqn2f} we find the bosonic chemical potential
in terms of the densities
\bea
\mu_B&=&g_{DD} n_2+g_{AD}(n_1-n_2)\label{eqmub1}
\eea
which is a result at linear order in density for the bosonic chemical potential $\mu_B$ defined in Eq.\ref{eqdefmuB}.
\section{Density expansion : next order}\label{densepx2}
 In order to get a diagram for $S_1$ at next order in $k_0$, we include the possibility to have a hole
 line (therefore going backward in a graph) in addition to one $\Delta$ and a $\Delta^*$. Such graphs are shown in Fig.\ref{fig2}(b). The arguments yielding to these graphs are given in Appendix 
 \ref{apps1b}.
\bea
S_1^{(b)}(k)&=&|\Delta|^2 \sum_p T_3^{1}(p,k;p)T_2(p-k) T_3^{1}(k,p;p)G_1^0(p)\nonumber\\
\eea
In order to calculate $\delta n_1^{(b)}=-\sum_k (G_1^0(k))^2 S_1^{(b)}(k)$, we need to perform two summations on the frequencies $\omega_k$ and $\omega_p$. In the summation on $\omega_p$, we use  the fact that $T_3^1(p,k;p)$ ($=T_3^1(k,p;p)$)
(this is quite easily seen by writing down the diagrammatic expansion where
one sees that $p$ always appears with a plus sign) and $T_2(p-k)$ are analytical for $\mathcal{R}e(i\omega_p)<0$  and we perform
the integration like in Eq.\ref{eqdnk1a}. By doing so, due to  
the analytical properties of $T_3^{1}(p,k;p)$ and $T_2(p-k)$ as a function of the frequency $i\omega_p$, we close the integration contour in $\mathcal{R}e(i\omega_p)<0$ and we get a contribution of the pole of $G_1^0(p)$ for
 $|{\bf p}|<k_F$ and in this way we find that $p$ is "on the shell" ($p=\{{\bf p},({\bf p}^2-k_0^2)/2 m_1\}$).
The summation on $\omega_k$ is also easily performed if we use the fact that $T_{2}(p-k)$ are analytical for $\mathcal{R}e(i\omega_k)>0$. Like in Eq.\ref{eqdnk1a}, we get two terms due to
the presence of a double pole in $(G_1^0(k))^2$, and $k$ is on "the shell"  
($k=\{{\bf k},({\bf k}^2-k_0^2)/2 m_1\}$). In the term with the derivative of the product
$T_3^1(p,k;p)^2 T_2(k-p)$, we see easily that the derivative of $T_3^1$ will give a finite term
in the low density limit (as can be see easily using for instance the integral equation for $T_3^1$).
$T_2(k-p)$ can be written as the product of the bosonic propagator which is singular at low energy and wave vector times a regular function at low energy. The derivative of $T_2(k-p)$ at low energy thus gives

\bea
\frac{\partial}{\partial i\omega_k}\left[
\frac{Z}{
\frac{{\bf p}^2-k_0^2}{2 m_1} -i\omega_k-\frac{({\bf p}-{\bf k})^2}{2 m_D}
}
\right]_{i\omega_k=\frac{({\bf k}^2-k_0^2)}{2 m_1}}\nonumber\\
=
\frac{Z}{\left[
\frac{{\bf p}^2-{\bf k}^2}{2 m_1} -i\omega_k-\frac{({\bf p}-{\bf k})^2}{2 m_D}
\right]^2}
\eea
We can now perform the integrations on the wave vectors ${\bf p}$ and ${\bf k}$. 
Like in Appendix \ref{appdt31} , it can be shown that the low wave vector domain is dominant and we get for the correction to the density of particles $1$
\bea
\delta n_1^{(b)}&=&
-N_0\left(g_{AD}\frac{|\Delta|^2}{Z}\right)\left(
k_0 \,a_{AD}\right)(C_{b}-C_{a,2})
\nonumber\\
\label{eqdn1b}
\eea
where the constant $C_b$ is defined by
\bea
C_b&=&4\pi \int\! \frac{d^3{\bf x}}{(2\pi)^3}
\frac{
\Theta(1-|{\bf x}|)(r+1)
}
{\left(
r(1-{\bf x}^2)+({\bf u}-{\bf x})^2
\right)
}
\label{eqCb}
\eea
where ${\bf u}$ is any unitary vector.
Finally, the total density of fermions of species $1$ is given in terms of $\Delta$ and $k_0$ ($\equiv
\sqrt{2 m_1 \mu_1}$) by the expansion

\bea
n_1 &=&
\frac{|\Delta|^2}{Z}
+\frac{k_0^3}{6\pi^2}\nonumber\\
&&-N_0\left(g_{AD}\frac{|\Delta|^2}{Z}\right)\left(
1+k_0 a_{AD}\,C(m_D/m_1)
\right)
\nonumber\\
&&+ \cdots\label{eqn1f}
\eea
where the constant $C(m_D/m_1)=C_{a,1}+C_b$. After calculation of the required integrals,
one finds that $C$ is given in terms of the function $f((m_D-m_1)/(m_D+m_1))$ given by Albus {\it et al.}
\cite{albusetal}. Specifically we find $\pi \,C(m_D/m_1)= (4/3) f(\delta)$,
where we have defined the dimensionless parameter
\bea
\delta\equiv \frac{(m_D-m_1)}{(m_D+m_1)}=\frac{m_2}{2\, m_1 +m_2}
\eea
 and we have
\bea
f(\delta)&=&\frac{3}{4}\frac{(\delta-1)}{\delta}+\frac{3\,(1+\delta)^2 (1-\delta)}{8\delta^2}\ln(\frac{1+\delta}{1-\delta})
\eea
For the particles $2$, we find naturaly that there is no contribution of diagrams similar
to the ones shown in Fig.\ref{fig2}(b) (but with the roles of particles $1$ and $2$ exchanged) because there is no hole for particles $2$ at lowest order.
The first term in the {\it r.h.s.} of Eq.\ref{eqn1f} is $n_2$ at the order we consider, due to Eq.\ref{eqn2f}. 
Moreover we can rewrite Eq.\ref{eqn1f} by considering it as a Taylor expansion, after replacing
$C$ by $4/(3\pi)f$
\bea
n_1-n_2&=&\frac
{
\left(
2 m_1\left[\mu_1 -g_{AD} n_2 (1+\frac{4}{3\pi}k_F a_{AD}\,f(\delta))\right]
\right)^{3/2}
}
{6\pi^2}\nonumber\\
\label{eqn1n2}
\eea
where we ignored terms of order $(density)^{5/3}$. At the order we consider we have replaced $|\Delta|^2 /Z$ by $n_2$ in the {\it r.h.s.} of the equation, and 
$k_0$ by Fermi wave vector $k_F$ in the term in factor of $f$. The Fermi wave vector is defined in terms of the density of fermions in excess in the usual way
\bea
k_F&\equiv &(6\pi^2 (n_1-n_2))^{1/3}
\eea
Eq.\ref{eqn1n2} is easily inverted and we find
\bea
\mu_1&=&\frac{k_F^2}{2\,m_1}
+g_{AD} n_2 \left(1+\frac{4\,k_F a_{AD}}{3\,\pi}
f(\delta)\right)\label{eqmu1f}
\eea

We can now easily calculate $\mu_2$ using Eq.\ref{eqmu1f} and \ref{eqmub1}. Indeed, we can follow \cite{albusetal} and use that $\mu_1=(\partial(E/V)/\partial n_1)_{n_2}$, where $E$ is the ground state energy in a volume $V$. We integrate Eq.\ref{eqmu1f} as a function of $n_1$ for $n_1\leq n_2$, keeping $n_2$ fixed. We get
\bea
\frac{E}{V}&=&E_B(n_2)+\int_{n_2}^{n_1}\!dn'_1\,\mu_1(n'_1,n_2)\label{eqEV1}
\eea 
where $E_B(n_2)$ is a constant of integration. We see from this equation it is the ground state energy (per unit volume) of the system with an equal number of particles $1$ and $2$ ("unpolarized" situation). We know from \cite{xlrclhy} it is given by the Lee-Huang-Yang (LHY) \cite{lhy} expression.
Inserting Eq.\ref{eqmu1f} into Eq.\ref{eqEV1}, we get, using the mean field term like term
of the LHY expression for $E_B(n_2)$
(the LHY beyond mean-field correction is of order $(density)^{5/2}$ which we neglect at the order we consider)
\bea
\frac{E}{V}&=&-|E_b| n_2 +\frac{1}{2}g_{DD}(n_2)^2
+\frac{3}{10}\frac{k_F^5}{m_1}\nonumber\\
&&+g_{AD}(n_1-n_2) n_2 \left(1+\frac{k_F a_{AD}}{\pi}
f(\delta)\right)\nonumber\\
\label{eqEV}
\eea
which is an expansion up to terms of order $(density)^{7/3}$.
From this we can take the derivative with respect to $n_2$ and we find for the chemical
potential of particules $2$ 
\bea
\mu_2&=&-|E_b|-\frac{k_F^2}{2 m_1}+g_{DD}n_2\nonumber\\
&&+g_{AD}(n_1-2 n_2)\nonumber\\
&&+g_{AD}\left(\frac{ k_F\, a_{AD}f(\delta)}{3\pi}(3 n_1-7n_2)\right)
\nonumber\\\label{eqmu2}
\eea
where we have included  terms up to order $(density)^{4/3}$.
The bosonic chemical potential $\mu_B \equiv \mu_1+\mu_2+|E_b|$ is then from
Eq.\ref{eqmu1f} and \ref{eqmu2}
\bea
\mu_B&=&g_{DD}n_2\nonumber\\
&&+g_{AD}(n_1- n_2)\left(1+\frac{ k_F\,a_{AD}f(\delta)}{\pi} \right)
\nonumber\label{eqmubf}\\
\eea
Looking at Eq.\ref{eqmu1f} and \ref{eqmubf}, we find that we recover precisely at this order the result of Albus {\it et al.} \cite{albusetal} Eqs. $76$ and $67$ respectively
if we consider that there are $n_2$ dimers considered as pointlike bosons (mass $m_D$) and $n_1-n_2$ fermions (mass $m_1$). The boson-boson interaction  is described by the scattering length $a_{DD}$
and the fermion-bosons interaction is characterized by the scattering length $a_{AD}$.
\section{Impurity problem and one-particle gap}\label{gapimp}
For a system with $N_1$ particles $1$ and $N_2$ particles $2$, we denote the groundstate energy $E(N_1,N_2)$. A single particle $2$ (an "impurity") immersed in a Fermi Sea of particles $1$ has an excitation energy
\bea
E(N_1,1)-E(N_1,0)&=&\mu_2(n_1,n_2=0)
\eea
In the BEC limit, this problem  has been the subject of many studies recently both experimentally \cite{expimp} and theoretically \cite{theoimp}.
From Eq.\ref{eqmu2} with $n_2=0$, we find
\bea
\mu_2^{imp}&=&-|E_b|-\frac{k_F^2}{2 m_1}\nonumber\\
&&+g_{AD}\,n_1\,\left(1+\frac{ k_F\,a_{AD}f(\delta)}{\pi}\right)
\eea
where $k_F\equiv(6\pi^2 n_1)^{1/3}$.
This result, and also a next-order correction, was found by Mora and Chevy in Ref.\cite{theoimp}.

The one-particle gap $\Delta_{gap}$ is related to the ground-state energy via
\cite{gps},\cite{ringschuck}
\bea
\Delta_{gap}&=&\frac{1}{2}\left(2\,E(N+1,N)-E(N+1,N+1)-E(N,N)\right)\nonumber\\
&=&\frac{1}{2}\left(\mu_1-\mu_2\right)
\eea
From Eq.\ref{eqmu1f},\ref{eqmu2}, we find
\bea
\Delta_{gap}&=&\frac{1}{2}\left(|E_b|+(2 g_{AD}-g_{DD})n_2\right)
\eea
a result given in Eq.$(40)$ (for particles of equal masses) of Ref.\cite{gps}.
\section{Comments and conclusion}\label{conclusion}

We have presented in this work a low density expansion of the equation of state 
of a condensate of composite bosons made of two fermions of two different species, in the case where one of the two species is in excess. We have shown results of the expansion for the chemical potentials up to order $(density)^{4/3}$. However there must be terms of the type of the LHY correction of order $(density)^{3/2}$
(more precisely of the type $n_2^{3/2}$ and most probably of type $n_1 n_2^{1/2}$). 
In order to see the quantitative importance of the LHY correction and the correction to
$\mu_B$ given in Eq.\ref{eqmubf}, we calculate the ratio $R$ between these two terms \bea
R&=&\frac{ g_{DD}n_2 \frac{32}{3\sqrt{\pi}}\left(
n_2 a_{DD}^3\right)^{1/2}}
{g_{AD}(n_1-n_2)k_F a_{AD}\frac{f(\delta)}{\pi}}\nonumber\\
&=&0.258\left( \frac{n_2}{n_F} \right)^{4/3}\sqrt{k_{F,b}\,a}
\eea
with $n_F\equiv n_1-n_2$ and $k_{F,b}\equiv (6\pi^2 n_2)^{1/3}$.
 We have taken $m_1=m_2$, $a_{DD}/a=0.6$ \cite{add}\cite{4parpra}, $a_{AD}/a=1.18$ \cite{aad} in order to get the numerical prefactor. In principle, for fermions densities of the same order of magnitude,
 the ratio $n_2/n_F$ is of order $1$, and the ratio $R$ tends to $0$ in the dilute limite
 $\sqrt{k_{F,b}\,a}\to 0$. However, in the experiment of Ref.\cite{YShinKetterle}, looking at Fig.$4$a, we estimate at the center of the trap $n_2/n_F\approx 8$ and $k_{F,b}\,a\approx 1$. This gives $R\approx 4.1$. Close to the edge of the bosonic cloud,
 if we estimate that the dimer density is divided by $2$, this gives $\sqrt{k_{F,b}\,a}\approx 0.9$ and for a ratio $n_2/n_F\approx 4$, we find 
 $R\approx 1.46$. From this order of magnitude estimate, we see that including the corrections of Eqs.\ref{eqmu1f},\ref{eqmu2} and \ref{eqmubf} for the equations of state (in addition to the LHY correction as it was done in Ref.\cite{YShinKetterle}) should improve quantitatively the analysis of the experiment of Ref.\cite{YShinKetterle}.

{\bf Acknowledgments:} We thank F. Chevy, R. Combescot, C. Mora, P. Pieri and G. C. Strinati for interesting discussions. F. A. thanks the CNRS for financial support.

\appendix\label{details}
\section{Corrections to $T_{3}^1$}\label{appdt31}
We want first to determine  at lowest order in the densities the difference 
$T_3^{1}(k, k;k)-T_{3,v}^{1}(0,0;0)$ where $k=\{{\bf k},({\bf k}^2 -k_0^2)/(2 m_1)\}$
and $|{\bf k}|\ll a^{-1}$. $T_{3,v}^{1}$ is the dimer-atom $1$ $T$-matrix {\it in vacuum}.
It is obtained from $T_3^{1}$ by setting $\mu_1$ to $0$ and $\mu_2$ to $-|E_b|$.
One can define in the same way $T_{2,v}$, $G_{1}^{v}$ and $G_{2}^v$.
We write $T_3^{1}(k, k;k)-T_{3,v}^{1}(0,0;0)=(T_3^{1}(k, k;k)-T_{3,v}^{1}(k, k;k))
+(T_{3,v}^{1}(k, k;k)-T_{3,v}^{1}(0,0;0))$, and we calculate the two differences between parentheses separately.
\subsection{$T_3^{1}(k, k;k)-T_{3,v}^{1}(k, k;k)$}\label{subappA1}
\begin{figure}[h]
\begin{center}
\includegraphics[width=8cm]{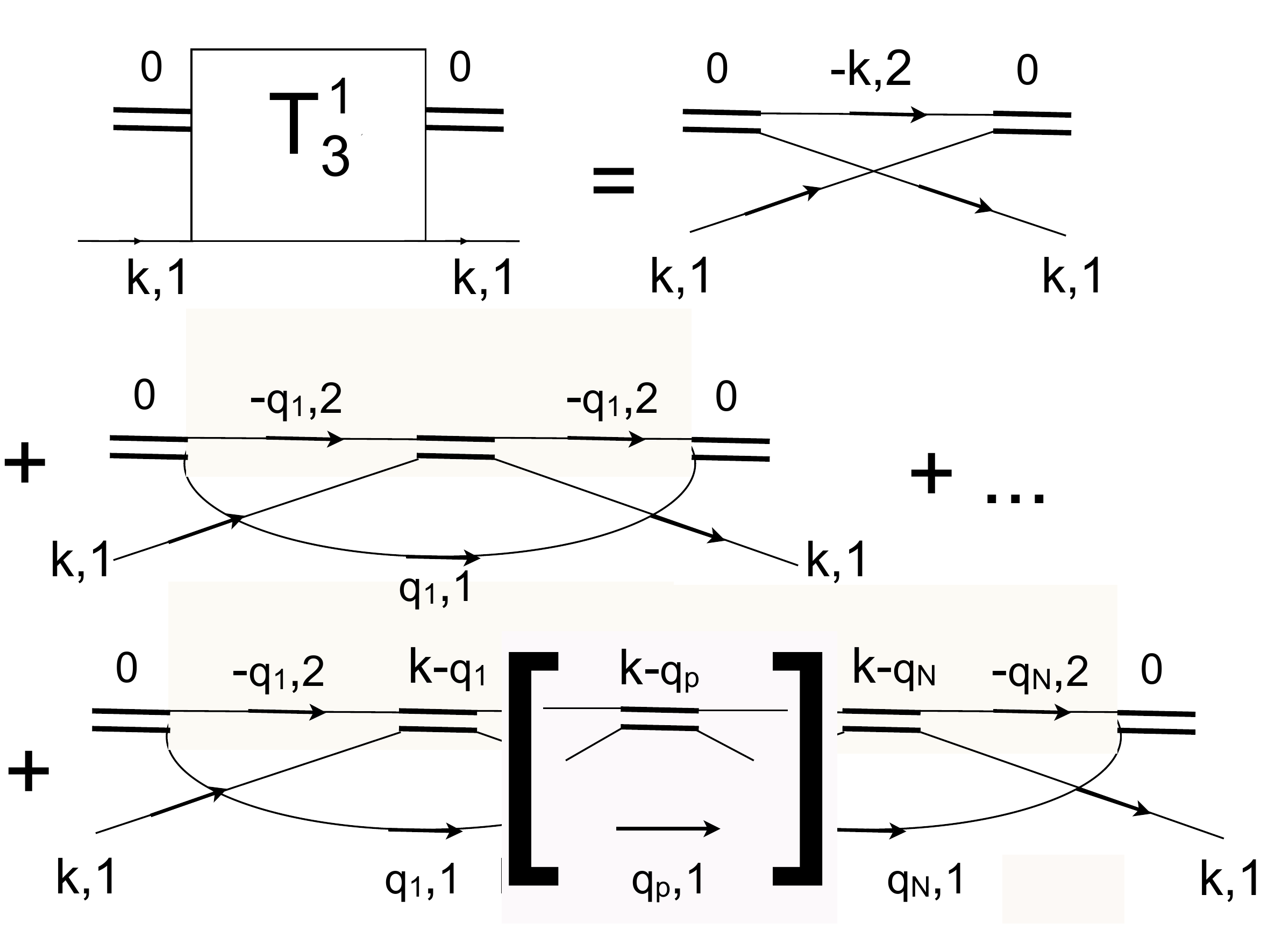}
\caption{The diagrammatic expansion of $T_{3}^1(k,k;k)$.}
\label{figappA}
\end{center}
\end{figure}
In order to calculate the first difference, we note that it comes from the fact that the fermionic propagators and the dimer propagator are different in vacuum and with
$\mu_1\neq 0$ and $\mu_2 \neq -|E_b|$. The principle of the method is therefore the following : we write $\Delta G_{1,2}(k)=G_{1,2}^{0}(k)-G_{1,2}^{v}(k)$, $\Delta T_2(K)=T_2(K)-T_{2,v}(K)$ and  we expand the diagrammatic series for $T_3^1$ in powers of 
$\Delta G_{1,2}$ and $\Delta T_2$.
If we look at the diagrams for $T_3^1(k,k;k)$ with $N$ dimer propagators $T_2$, we see that $G_{1}^{0}$ appear $N$ times in $G^1_0(q_p)$ for $p=1,\cdots N$ (see Fig.\ref{figappA}). If we write $G_1^0(q_p)=G_1^v(q_p)+\Delta G_1(q_p)$, the term
of order $\Delta G_1(q_p)$ will be multiplied by $T_{2,v}(k-q_p)$, by all the diagrams of 
$T_{3,v}^1(k,k;k)$ on his left and by all the diagrams of $T_{3,v}^1(k,k;k)$ on his right. However those diagrams are Feynman diagrams for $T_{3,v}^1(k,q;k)$ and $T_{3,v}^1(q,k;k)$ respectively. Therefore when we resum all the diagrams with any number of $T_2$ we have 
\bea
&&\sum_q T_{3,v}^1(k,q;k)T_{2,v}(k-q)\Delta G_1(q) T_{3,v}^1(q,k;k)\label{eqA1}
\eea
where we have taken the atom-atom and atom-dimer $T$ matrices in vacuum at lowest order.
By looking at the diagrammatic expansion of $T_{3,v}^1(k,q;k)$
($=T_{3,v}^1(q,k;k)$), we find that  it is analytic for $\mathcal{R}e(i\omega_q)>0$ ($q=\{{\bf q}, i\omega_q\}$ appears always with a minus sign in a particle $2$ propagator $G_2$). For the same reason $T_2(k-q)$ is analytic for $\mathcal{R}e(i\omega_q)>0$. 
In the zero temperature limit, we can therefore transform the frequency summation in Eq.\ref{eqA1} in an 
integral along the imaginary axis. Then we deform this contour in the half-plane 
$\mathcal{R}e(i\omega_q)>0$.  We get the two contributions of the poles of $\Delta G_1(q)$
at the frequencies $({\bf q}^2-k_0^2)/2 m_1$ and ${\bf q}^2/2 m_1$ ("on the shell") for
$G^1_0(q)$ and $G^1_v(q)$ respectively
\begin{equation}
\begin{split}
\int\! \frac{d^3{\bf q}}{(2\pi)^3}
[(T_{3,v}^1(k,q;k))^2 T_{2,v}(k-q)\Theta(|{\bf q}|-k_0)\\
-
(T_{3,v}^1(k,q;k))^2 T_{2,v}(k-q))]\label{eqdt31}
\end{split}
\end{equation}
where $q=\{{\bf q},({\bf q}^2-k_0^2)/2 m_1\}$ and $q=\{{\bf q},{\bf q}^2/2 m_1\}$
for the first and second terms respectively. The main contribution to the integral comes from small wavevectors $|{\bf q}|\ll a^{-1}$ as is explained below. In this case, we can replace $T_{3,v}$ by its low energy value $T_{3,v}(0,0;0)=Z^{-1} g_{AD}$ (the $Z$ factor comes from the two external dimer lines, each one giving a $\sqrt{Z}$ factor\cite{4parpra,levinsengurarie}).
For such small wave vectors, we can also
take $T_{2,v}(\{ {\bf K}, i\Omega_K \})\approx Z/(i\Omega_K -{\bf K}^2 /2 m_D)$ (a bosonic propagator multiplied by two $\sqrt{Z}$ factors for the incoming and outgoing dimer end). 
We therefore get at lowest order 
\bea
\begin{split}
&T_3^{1}(k, k;k)-T_{3,v}^{1}(k, k;k)=Z^{-1} (g_{AD})^2 \times\\
&\int\! \frac{d^3{\bf q}}{(2\pi)^3}
\left[
\frac{
\Theta(|{\bf q}|-k_0)
}
{
\frac{({\bf k}^2 -{\bf q}^2)}{2 m_1}-\frac{({\bf k} -{\bf q})^2}{2 m_D}
}
-
\frac{1}{\frac{({\bf k}^2 -k_0^2-{\bf q}^2)}{2 m_1}-\frac{({\bf k} -{\bf q})^2}{2 m_D}}
\right]
\end{split}\nonumber\\
\label{eqdt31f}
\eea
We note that this correction is of order $k_0$ (since $|{\bf k}|$ is of order $k_0$),
{\it i.e.} of order $(density)^{1/3}$, due to the "small denominators"
of order $k_0^2$.

We show now that the contribution to integral \ref{eqdt31} for wavevectors
 $|{\bf q}|>\Lambda$, where $\Lambda$ is a cut-off, with $k_0\ll \Lambda \ll a^{-1}$ is negligible.
 Indeed, in such a range of wavevectors, we can perform a Taylor expansion
 of the difference in Eq.\ref{eqdt31} with respect with the energy  variable 
 of $q$. The energy difference is of order $k_0^2$, and by close inspection of the high
 wavectors behavior of $T_3^1$ and $T_2$, we find a converging integral at infinity. 
 The low wave vector behavior ($\sim \Lambda$) is dominated by the small denominators of the $T_2$'s and this gives a contribution of order $k_0 ^2 \Lambda$.
 Compare to the term of Eq.\ref{eqdt31f}, this is negligible if $k_0 \Lambda a^2 \ll 1$.
 This is possible if $\Lambda= (k_0/a)^{1/2}$ for instance.
 As a conclusion, the high wavevector domain gives a negligible contribution.
 
 We can make the same sort of analysis with $\Delta T_2$. In this case, we would find 
 that the characteristic low wave vector is of order the square root of the density
 (this is obtained by comparing $\mu_B$ and $R$ to ${\bf K}^2$ in the denominator of
 Eq.\ref{eqT2}). This gives a contribution of the order of $(density)^{1/2}$
 instead of $k_0$ which is of the order of $(density)^{1/3}$. Therefore it is negligible.
 
 We can also consider the effect of $\Delta G_2$. There is no "small denominator"
 in this case, since one has to compare the kinetic energy ${\bf q}^2 /2 m_2$ to the
 (large) chemical potential $\mu_2\approx -|E_b|$. As a result, we find the correction is of order the chemical potential correction  $\mu_2 +|E_b|$, which is of order $k_0^2$. Therefore it is negligible here.

\subsection{$T_{3,v}^{1}(k, k;k)-T_{3,v}^{1}(0,0;0)$}\label{appdt312}
In order to calculate this difference, we note that in the diagrammatic expansion of
$T_{3,v}^{1}(k, k;k)$, the $4$-wavevector $k$ appears only in $G_2^{v}(k-q_p-q_{p+1})$
and in $T_2(k-k_p)$ (not in the $G_1^v$'s). We define now 
$\Delta T_2(k-q_p)=T_2(k-q_p)-T_2(-q_p)$ and similarly for $\Delta G_2$.

The corrections of order $\Delta G_2$ gives a negligible term again, for the same reason as given in subsection \ref{subappA1}.

The correction of order $\Delta T_2$ can be analyzed like the correction of order
$\Delta G_1$ above. Indeed, we can permute the roles of $T_2(k-q_p)$ and $G_1(q_p)$
in the argument. In this way, we find

\bea
&&\sum_q T_{3,v}^1(0,q;0)\Delta T_{2,v}(k-q) G_1^v(q) T_{3,v}^1(q,0;0)
\eea
The frequency summation is done as above and we get the contribution "on the shell"

\begin{equation}
\begin{split}
\int\! \frac{d^3{\bf q}}{(2\pi)^3}
(T_{3,v}^1(0,\{{\bf q},\frac{{\bf q}^2}{2 m_1}\};0))^2 
&(T_{2,v}(\{{\bf k}-{\bf q},\frac{ {\bf k}^2-k_0^2-{\bf q}^2}{2 m_1 }\})\label{eqdt32}\\
-
&T_{2,v}(\{-{\bf q},-\frac{{\bf q}^2}{2 m_1 }\})
\end{split}
\end{equation}
We can show again that the low wave vector domain gives the main contribution, and replace the atom-dimer $T$ matrix by $Z^{-1} g_{AD}$ and the $T_2$'s by their bosonic expressions. In this way we get at lowest order

\bea
\begin{split}
&T_{3,v}^{1}(k, k;k)-Z^{-1}g_{AD}=Z^{-1} (g_{AD})^2 \times\\
&\int\! \frac{d^3{\bf q}}{(2\pi)^3}
\left[
\frac{1}
{
\frac{({\bf k}^2 -k_0^2-{\bf q}^2)}{2 m_1}-\frac{({\bf k}-{\bf q})^2}{2 m_{D}}
}
+
\frac{1}{\frac{{\bf q}^2}{2 m_{AD}}}
\right]\\
\label{eqdt32f}
\end{split}
\eea

Collecting Eq.\ref{eqdt31f} and \ref{eqdt32f}, we get

\bea
&T_{3}^{1}(k, k;k)-Z^{-1}g_{AD}=\nonumber\\
&Z^{-1} (g_{AD})^2 (k_0) m_{AD}
F_{a,1}\left(\frac{|{\bf k}|}{k_0},\frac{m_D}{m_1}\right)\label{eqdt3f}
\eea
where the dimensionless function $F_{a,1}$ is defined by 
\bea
F_{a,1}(x,r)&=&2\int\! \frac{d^3{\bf y}}{(2\pi)^3}
\left[
\frac{
\Theta(|{\bf y}|-1)(r+1)
}
{
r({\bf x}^2 -{\bf y}^2)-({\bf x}-{\bf y})^2
}
+
\frac{1}{{\bf y}^2}
\right]\nonumber\\
\label{eqFa1}
\eea

\section{Evaluation of $\frac{d}{dz} T_3^{1}(k,k;k)$}\label{appddt31}
In order to evaluate this derivative, we note again that $k$ appears in the diagrammatic expansion
of $T_3^1 (k, k;k)$ in the propagators of the particles  $2$ $G_2(k-q_p-q_{p+1})$ and in the
dimer propagators $T_2(k-q_p)$. 
If we take the derivative of $T_2(k-q)$ with respect to the frequency of $k$, which we denote $z$ here, we get similarly
to \ref{appdt312}
\bea
&&\sum_q T_{3}^1(k,q;k)\frac{d}{d z}T_{2}(k-q) G_1^0(q) T_{3}^1(q,k;k)
\eea
We can perform the frequency summation as before, which has the effect to take the 
frequency "on the shell" ($i\omega_q=({\bf q}^2 -k_0^2)/2 m_1$) and $|{\bf q}|>k_0$. 
Again, the low wave-vector domain gives the main contribution. We can safely replaces the
$T_3^1$'s by $Z^{-1}g_{AD}$ and $T_2$ by the bosonic expression. We get at lowest order
(using $k=\{{\bf k},({\bf k}^2 -k_0^2)/2m_1\}$)
\bea
\frac{d}{dz} T_3^{1}(k,k;k)&=&-Z^{-1}(g_{AD}m_{AD})^2\frac{1}{k_0}F_{a,2}\left(\frac{|{\bf k}|}{k_0},\frac{m_D}{m_1}\right)\nonumber\\
\label{eqddt3f}
\eea
where
\bea
F_{a,2}(x,r)&=&4\int\! \frac{d^3{\bf y}}{(2\pi)^3}
\frac{
\Theta(|{\bf y}|-1)(r+1)^2
}
{\left(
r({\bf x}^2 -{\bf y}^2)-({\bf x}-{\bf y})^2
\right)^2}
\nonumber\\
\label{eqFa2}
\eea
\section{Calculation of $n_2^{(a)}$}\label{appn2}
We proceed like for the calculation of correction to \mbox{$T_3$ :} for each propagator
($G_1^0$, $G_2^0$ or $T_2$) we write it as the sum of its value in vacuum plus a difference. In this way, we are left with most of the propagators in vacuum.
Therefore we start with all diagrams in vacuum for $S_2^{(a)}$. The calculation of the contribution to $n_2$ is then similar to Ref.\cite{xlrclhy}, due to the analytical properties
of $T_{3,v}^2(k,k;k)$ as a function of the frequency. We find
$
n_2 = \frac{|\Delta|^2}{Z}
$. In the course of the calculation, we have used $2\mu\approx -|E_b|$ and we have neglected terms of the order of the density.

We turn now to corrections to propagators.
\subsection{Correction of order $\Delta G_1$}
We first consider corrections of order $\Delta G_1$. Looking at the diagrams for 
$S_2^{(a)}$, they look like Fig.\ref{figappA}, but with the roles of particles $1$ and $2$ exchanged. We see that $G_1$ appears in four places : {\it i)} $G_1^0(-q_1)$,
{\it ii)} $G_1^0(-q_N)$, {\it iii)} $G_1^0(k-q_p-q_{p+1})$ and {\it iv)} $G_1^0(-k)$.

In case  {\it i)}, we find the contribution to $n_2$
\bea
|\Delta|^2\sum_{k,q_1}(G_{2}^{v}(k))^2 \Delta G_1(-q_1) T_2(k-q_1)G_{2}^v(q_1)
T_{3,v}^2(q_1,k;k)\nonumber
\eea
But in the diagrammatic expansion of $T_{3,v}^2(q_1,k;k)$, the variable $k$ only appears with a plus sign. This means $T_{3,v}^2(q_1,k;k)$ is analytical for 
$\mathcal{R}e(i\omega_k)<0$. This is also true for $T_2(k-q_1)$, therefore, when we sum on $\omega_k$, we find zero. The case {\it ii)} is similar if we replace $q_1$ by $q_N$, and we also find a result equal to zero.

The case {\it iii)} is different. We find,  similarly to what was found in \ref{subappA1}, the following contribution to the density of particles $2$
\bea
\sum_{k,q,q'}
(G_{2}^v(k))^2 
T_{3,v}^2(k,q;k)T_2(k-q)G_{2}^{v}(q)\nonumber\\
\Delta G_1(k-q-q')
T_{3,v}^2(q',k;k)T_2(k-q')G_{2}^{v}(q')\label{eqiii}
\eea
All the integrand except $\Delta G_1(k-q-q')$ is analytical for $\mathcal{R}e(i\omega_k)<0$. As usual by deforming the contour of integration in the half plane  $\mathcal{R}e(i\omega_k)<0$, we get a pole of $G_1^0(k-q-q')$ for
$i\omega_k=i\omega_q+i\omega_{q'}+(({\bf k}-{\bf q}-{\bf q'})^2 -k_0^2)/(2 m_1)$
if $|{\bf k}-{\bf q}-{\bf q'}|<k_0$. The difference $(({\bf k}-{\bf q}-{\bf q'})^2 -k_0^2)/(2 m_1)$ is small compared to the binding energy $|E_b|$, which is the energy scale of $T_{3,v}^1$. Therefore we can replace at lowest order $i\omega_k$ by $i\omega_q+i\omega_{q'}$ in Eq.\ref{eqiii}. In the same manner, since $|{\bf k}-{\bf q}-{\bf q'}|<k_0$ is much smaller than  $a^{-1}$ a typical wave vector scale for $T_{3,v}^1$, we can replace ${\bf k}$ by ${\bf q}+{\bf q}'$ in Eq.\ref{eqiii}. This means that we can replace everywhere, at lowest order,  the $4$-momentum $k$ by $q+q'$ in Eq.\ref{eqiii}. The integrand is then independant of ${\bf k}$ and we get a phase space factor $k_0^3/(6\pi^2)$. The remaining summations on $q$ and $q'$ give the factor
\bea
\sum_{q,q'}
(G_{2}^{v}(q+q'))^2 
T_{3,v}^2(q+q',q;q+q')T_2(q')G_{2}^{v}(q)\nonumber\\
T_{3,v}^2(q',q+q';q+q')T_2(q)G_{2}^{v}(q')\label{eqiiibis}
\eea
We can perform the summation on the frequencies as before (only the "Born" contributions to $T_{3,v}^2$ contribute) and we get the two contributions "on the shell" $i\omega_q =-{\bf q}^2 /(2 m_1)$ and $i\omega_{q'} =-{\bf q'}^2 /(2 m_1)$
\bea
 \int \!  \frac{d^3{\bf q}d^3{\bf q'}}{(2\pi)^6}
 T_2({\bf q},-\frac{{\bf q}^2}{2 m_1})
 G_{2}^{v}({\bf q},-\frac{{\bf q}^2}{2 m_1})
T_2({\bf q'},-\frac{{\bf q'}^2}{2 m_1})\nonumber\\
 G_{2}^{v}({\bf q'},-\frac{{\bf q'}^2}{2 m_1})
 \left(
 G_{2}^{v}({\bf q}+{\bf q'},-\frac{{\bf q}^2+{\bf q'}^2}{2 m_1})
 \right)^2\label{eqC}
\eea
This integral is convergent close to ${\bf q}={\bf 0}$  and ${\bf q'}={\bf 0}$. It is also easily shown that it converges when the wave vectors tend to infinity. The integral \ref{eqC} is therefore a constant, independant of $k_0$.
As a conclusion, this case {\it iii)} gives a contribution to $n_2$ of order $|\Delta|^2 (k_0)^3=(density)^2$ and we can safely neglect it.

Finally we consider the case  {\it iv)}. This is given by
\bea
|\Delta|^2\sum_k (G_{2}^{v}(k))^2 \Delta G_1(-k)
\eea
The frequency integration is easily done and we find
\bea
&
|\Delta|^2 \int \!  \frac{d^3{\bf k}}{(2\pi)^3}
\left[
\frac{\Theta(k-k_0)}
{\left(
\frac{{\bf k}^2+a^{-2}}{2 m_r}-\mu_1
\right)^2
}
-\frac{1}{\left(\frac{{\bf k}^2+a^{-2}}{2 m_r}\right)^2}
\right]\nonumber\\
\approx &\mu_1 |\Delta|^2 2 \int \!  \frac{d^3{\bf k}}{(2\pi)^3}
\frac{1}{\left(\frac{{\bf k}^2+a^{-2}}{2 m_r}\right)^3}
\eea
Performing the integral on ${\bf k}$, we find the contribution to $n_2$
\bea
\mu_1 |\Delta|^2 \pi (m_r a)^3\label{eqdn21}
\eea
\subsection{Correction of order $\Delta G_2$}
Next, we study the contributions of order $\Delta G_2$.
In a diagram for $(G_2^0(k))^2 S_2^{(a)}(k)$, a particle $2$ propagator can be
{\it i)} in the two external propagators $G_2^0(k)$ or
{\it ii)} in the dimer-(fermion $2$) $T$-matrix $T_{3}^2(k,k;k)$.
In case {\it i)} the correction to $n_2$ of order $\Delta G_2$ is given by
\bea
-2 |\Delta|^2 \sum_k G_{2}^{v}(k) \Delta G_2 (k) T_{3,v}^2(k,k;k)
\eea
We can then write $T_{3,v}^2$ as the sum of $-G_{1}^{v}(-k)$, which has a pole for
$-{\bf k}^2/(2m_1)<0$, plus a function which is analytical for $\mathcal{R}e(i\omega_k)<0$
\cite{xlrclhy}. The frequency summation is then done as usual by deforming the contour of inegration in the half plane $\mathcal{R}e(i\omega_k)<0$. We get in this way, keeping terms of order $(density)^{2/3}|\Delta|^2$
\bea
-2(\mu_2+|E_b|)|\Delta|^2 \int \!  \frac{d^3{\bf k}}{(2\pi)^3}\frac{1}{(\frac{{\bf k}^2+a^{-2}}{2 m_r})}
\eea
But $\mu_2 +|E_b|=\mu_B-\mu_1=-\mu_1+O(density)$. And at lowest order we
can replace $-\mu_2$ by $\mu_1$. This gives the contribution
\bea
-\mu_1 |\Delta|^2 \pi (m_r a)^3\label{eqdn22}
\eea
The case ${\it ii)}$ is similar to what we discussed in \ref{subappA1}, except that the roles of particles $1$ and $2$ are exchanged. Therefore we find for the contribution to $n_2$
\bea
\sum_{k,q}(G_{2}^{v}(k))^2 T_{3,v}^2(k,q;k)  T_2(k-q) \Delta G_{2}(q)
T_{3,v}^2(q,k;k)\nonumber\\
\label{eqDeltaG2}
\eea
But all the functions are analytical for $\mathcal{R}e(i\omega_k)<0$. Therefore, we can perform the summation on $\omega_k$ by transforming it into an integral, and
deform the contour in the half-plane $\mathcal{R}e(i\omega_k)<0$. In this way, we find
that the sum of Eq.\ref{eqDeltaG2} is zero.

\subsection{Correction of order $\Delta T_2$}
We finish this appendix by adressing the contributions to $n_2$ of order $\Delta T_2$.
They are similar to the ones of order $\Delta G_2$, since for every dimer propagator
$T_2$ there is a fermion $2$ propagator $G_{2}$. We find
\bea
\sum_{k,q}(G_{2}^{v}(k))^2 T_{3,v}^2(k,q;k) \Delta T_2(k-q) G_{2}^{v}(q)
T_{3,v}^2(q,k;k)\nonumber\\
\label{eqDeltaT2}
\eea
Similarly to Eq.\ref{eqDeltaG2}, we find that this contribution is zero.
\subsection{Conclusion}
As a conclusion of this Appendix, we find that the two  corrections Eqs.\ref{eqdn21},\ref{eqdn22} of order
$|\Delta|^2 k_0^2$ exactly compensate, and we find $n_2=|\Delta|^2/Z$ and there is no correction up to order
$(density)^{5/3}$ included.
\section{Feynman diagrams of $S_1^{(b)}$}\label{apps1b}
\begin{figure}[h]
\begin{center}
\includegraphics[width=8cm]{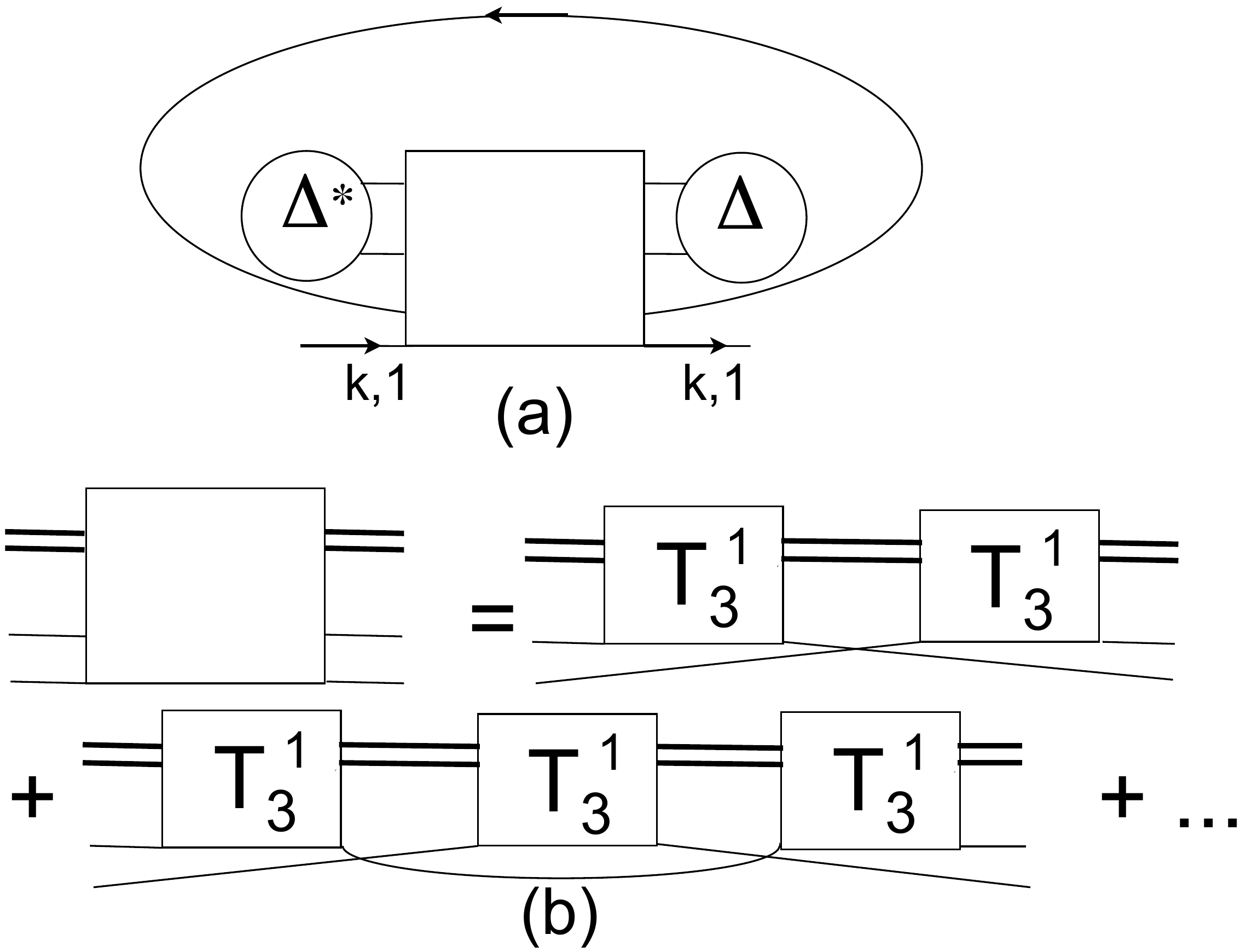}
\caption{(a) A general diagram for $S_1(k)$ with one hole line, one $\Delta$ and one $\Delta^*$. (b) Diagrammatic expansion of the new vertex. Only the first diagram contributes at lowest order (see text).}
\label{figappD}
\end{center}
\end{figure}
We explain here why the Feynman diagrams for the generalized self-energy $S_1(k)$
are given by Fig.\ref{fig2}(b) at the order we consider. We look for graphs with two anomalous self-energies (one $\Delta$, one $\Delta^*$) and one hole line. In general this can be drawn as shown in Fig.\ref{figappD}(a). The white box is a new vertex with one incoming dimer (made of a fermion of species $1$ and a fermion of species $2$)
, two incoming fermions of species $1$ and one outgoing dimer and two outgoing fermions of species $1$. This sort of vertex is very similar to the vertex $\Phi$ introduced in \cite{4parpra} with one incoming dimer and two incoming fermions, which {\it do not} interact first (this condition is automatically fulfilled here since the two incoming fermions are of the same species). In this case, one of the incoming fermions $1$ first interacts with the fermion $2$ of the incoming dimer. All the subsequent 
processes are resummed in the dimer-fermion $T$-matrix $T_3^1$ and we have one dimer and two fermions $1$ at the output of $T_3^1$. Then the other fermion $1$ interacts with the dimer, giving an other $T_3^1$, (see Fig.\ref{figappD}(b)) etc ...
Consider the first term with two $T_3^1$'s in Fig.\ref{figappD}(b). In order to get the generalized self-energy, we have to close a fermion $1$ loop.  It is then rather easy to see that in order to avoid double counting, we
must consider the diagram of Fig.\ref{fig2}(b). As is shown in Eq.\ref{eqdn1b}, this gives a contribution
to the density of fermions $1$ of order $|\Delta|^2 k_0^2$. This order of magnitude can be
estimated easily : the condition $|{\bf k}|< k_0$ gives a factor $k_0^3$, the hole line in $S_1$
gives an other factor $k_0^3$. But there is a large term of order $1/k_0^4$ coming from the derivative of $T_2(p-k)$. This is so because {\it both} $p$ {\it and} $k$ are at low energy and therefore 
there is a small energy denominator of order $k_0^2/(2 m_1)$.
If we consider now the other terms with three or more $T_3^1$, we will get the same phase space 
factor $(k_0)^6$, but as it is easily seen by writing the diagrams, the dimer propagators will never
be at low energy, and we will not get the large prefactor $1/(k_0)^4$. This concludes the justification
of retaining diagrams of Fig.\ref{fig2}(b).



\begin{thebibliography}{99}
\bibitem{YShinKetterle}Yong-il Shin {\it et al.} Phys. Rev. Lett. {\bf 101}, 070404 (2008)
\bibitem{becbcsmf} D. E. Seehy and L. Radzihovsky, Phys. Rev. Lett. {\bf 96}, 060401 (2006)
\bibitem{pilatigiorgini} S. Pilati and S. Giorgini Phys. Rev. Lett. {\bf 100}, 030401 (2008)
\bibitem{pieristrinati}P. Pieri and G. C. Strinati Phys. Rev. Lett. {\bf 96}, 150404 (2006)
\bibitem{iskinsademelo} M. Iskin and C. A. R. S\`a de Melo Phys. Rev. A {\bf 77}, 013625
(2008)
\bibitem{edtaylor}E. Taylor, A. Griffin and Y. Ohashi Phys. Rev. A {\bf 76}, 023614 (2007)
\bibitem{albusetal}A. P. Albus, S. A. Gardiner, F. Illuminati and M. Wilkens Phys. Rev. A {\bf 65}, 053607 (2002)
\bibitem{vivgior}L. Viverit and S. Giorgini Phys. Rev. A {\bf 66}, 063604 (2002)
\bibitem{AGD} A. A. Abrikosov, L. P. Gorkov and I. E. Dzyaloshinski,
{\it Methods of quantum field theory in statistical physics} (Dover, 1975).
\bibitem{gps} S. Giorgini, L. P. Pitaevskii and S. Stringari Rev. Mod. Phys. {\bf 80} 1215 (2008)
\bibitem{xlrclhy}X. Leyronas and R. Combescot Phys. Rev. Lett. {\bf 99}, 170402 (2007);
R. Combescot and X. Leyronas Phys. Rev. A {\bf 78}, 053621 (2008)
\bibitem{praclk}R. Combescot, X. Leyronas and M. Yu. Kagan Phys. Rev. A {\bf 73}, 023618 (2006)
\bibitem{expimp}G. Partridge {\it et al.} Science {\bf 311}, 503 (2006); 
M. Zwierlein {\it et al.} Science {\bf 311}, 492 (2006);
S. Nascimb\`ene {\it et al.} arXiv:0907.3032
\bibitem{theoimp} N. Prokof'ev and B. Svistunov Phys. Rev. B {\bf 77}, 020408(R) (2008);
R. Combescot, S. Giraud and X. Leyronas, Europhys. Lett. {\bf 88}, 60007 (2009);
C. Mora and F. Chevy Phys. Rev. A {\bf 80}, 033607 (2009); 
M. Punk, P. T. Dumitrescu and W. Zwerger Phys. Rev. A {\bf 80},053605 (2009) 
\bibitem{4parpra}  I.V. Brodsky, A. V. Klaptsov, M. Yu Kagan, R. Combescot and
X. Leyronas JETP Letters vol. 82, issue 5 , page 273-278 (2005); Phys. Rev. A {\bf 73}, 032724 (2006)
\bibitem{lhy} T. D. Lee and C. N. Yang, Phys. Rev. {\bf 105}, 1119 (1957);
T. D. Lee, K. Huang and C. N. Yang, Phys. Rev. {\bf 106}, 1135 (1957)
\bibitem{ringschuck} P. Ring and P. Schuck, {\it The Nuclear Many-Body Problem}
(Springer-Verlag, New York) 1980
\bibitem{add}D. S. Petrov, C. Salomon, and G. V. Shlyapnikov, Phys. Rev. Lett. {\bf 93},
\bibitem{aad}G. V. Skorniakov and K. A. Ter-Martirosian, Zh. Eksp. Teor. Fiz. {\bf 31}, 775 (1956) [Sov. Phys. JETP {\bf 4}, 648 (1957)]
\bibitem{levinsengurarie}J. Levinsen and V. Gurarie, Phys. Rev. A {\bf 73}, 053607 (2006)
\end{thebibliography}
\end{document}